\documentclass[11pt]{article}

\usepackage{hyperref}
\usepackage{amsmath}

\begin{document}

\title{Baryon chiral perturbation theory with virtual photons and leptons}

\author{Nopmanee Supanam,$^{a,b}$ Harold W. Fearing$^a$ and Yupeng
  Yan$^{b,c}$ \\
  \\ \vspace{6pt} $^a$TRIUMF,Vancouver, British Columbia, V6T 2A3, Canada
  \\ $^b$ Suranaree University of Technology,
 Nakhon Ratchasima,
  30000, Thailand
  \\ $^c$ Thailand Center of Excellence in Physics, \\
  Commission Higher on Education, Bangkok 10400, Thailand }

%  E-mail: \email{n\_supanam@yahoo.com, fearing@fearing.ca, yupeng@sut.ac.th}}

\maketitle

\begin{abstract}
  We construct the general pion-nucleon SU(2) Lagrangian
  including both virtual photons and leptons for relativistic baryon
  chiral perturbation theory up to fourth order. We include the light
  leptons as explicit dynamical degrees of freedom by introducing new
  building blocks which represent these leptons.
\end{abstract}

%\keywords{Chiral Lagrangians, Chiral perturbation theory, Electromagnetic and weak processes}

%\begin{document}
%
%\newenvironment{tablehere}
%  {\def\@captype{table}}
%  {}
\section{Introduction}

Chiral perturbation theory (ChPT) is an effective field theory of the
Standard Model at the low energies. The theory was first proposed by
Weinberg \cite{ref:Wei79} and formulated for light mesons by Gasser
and Leutwyler \cite{ref:GL84, ref:GL85} when external momenta are
small relative to the chiral symmetry-breaking scale, $\Lambda \sim 1$
GeV. Their formulation was first applied to the mesonic sector and
later extended to pion-nucleon interactions \cite{ref:GSS88}.

The general Lagrangian for the purely mesonic sector has been
constructed up to sixth order \cite{ref:GL84, ref:GL85, ref:FS96,
  ref:BCE00}. The inclusion of virtual photons in the mesonic
Lagrangian was first considered by Urech \cite{ref:Ure95}. He
calculated the divergences of the generating functional of ChPT
including virtual photons to one-loop and thus obtained the
counterterms necessary in the Lagrangian to cancel these
divergences. This Lagrangian was then used to calculate the one-loop
contribution to the meson masses in the isospin limit and thus obtain
the electromagnetic corrections to Dashen's theorem for $\Delta m_K^2
- \Delta m_\pi^2 $ \cite{ref:Ure95}.  Moreover, the Lagrangian also
has been used to compute the electromagnetic corrections to the S-wave
scattering lengths for the processes $\pi^0 \pi^0 \rightarrow \pi^0
\pi^0$ and $\pi^+ \pi^- \rightarrow \pi^0 \pi^0$ \cite{ref:KU98}. More
recently, both virtual leptons, as well as photons, have been included
in the effective mesonic Lagrangian making possible a complete
treatment of electromagnetic isospin-violating effects in semileptonic
weak processes \cite{ref:Kne+00} .  This modified Lagrangian was then
applied to the decays $\pi \rightarrow \ell \nu_\ell$ and $K
\rightarrow \ell \nu_\ell$ and to obtain the radiative corrections to
the theoretical decay rates \cite{ref:Kne+00}.

For the pion-nucleon sector, a difficulty comes from the fact that the
nucleon mass does not vanish in the chiral limit, which makes the
power counting scheme different from the meson sector. ChPT and the
power counting problem in the pion-nucleon sector were first
considered by Gasser, Sainio and \v{S}varc \cite{ref:GSS88}. Several
solutions have been proposed for this power counting problem including
heavy baryon chiral perturbation theory (HBChPT) \cite{ref:JM91} which
has been widely used, see e.g. \cite{ref:S03}. There have also been
various relativistic approaches suggested,
e.g. \cite{ref:BL99,ref:FGJS03}.  So far the complete effective
Lagrangian in the strong pion nucleon sector has been constructed up
to fourth order in the small momenta \cite{ref:FMS98, ref:Fet+00}.
The Lagrangian with virtual photons included has been developed by
M\"{u}ller and Mei{\ss}ner \cite{ref:MM99}. They construct the
Lagrangian for both relativistic and heavy baryon cases and apply
their result to calculations of the electromagnetic contributions to
the nucleon self energy, of the nucleon scalar form factor, and of the
$\pi^0$-nucleon S-wave scattering lengths.

The piece that is still missing is the pion-nucleon Lagrangian
involving virtual leptons. The aim of this work is thus to construct
the effective pion-nucleon Lagrangian including both virtual photons
and leptons.  Such a Lagrangian is important and necessary for the
full treatment of electromagnetic corrections to the weak processes
involving pions and nucleons, analogous to the treatments in the meson
sector of Ref. \cite{ref:Ure95, ref:Kne+00}. The simplest application
of our new Lagrangian is to radiative corrections to neutron beta
decay (or muon capture) which we will discuss in a subsequent paper.

In Sec. 2, we introduce the new building blocks related to the
leptonic current and in Sec. 3 we obtain the effective chiral
pion-nucleon Lagrangian with the inclusion of virtual photons and
leptons up to fourth order.

\section{Building blocks}

In $\mathrm{SU(2)}$ in the chiral limit where the up and down quark
masses go to zero the QCD Lagrangian has a $\mathrm{SU(2)}_R \times
\mathrm{SU(2)}_L$ chiral symmetry which is spontaneously broken to its
vectorial subgroup $\mathrm{SU(2)}_V$. As a result, the pion fields
can be described as Goldstone fields $\phi_i (x)$
\begin{equation}
\phi (x) = \sum_{i=1}^3 \tau_i \phi_i (x) \equiv \left(
    \begin{array}{c c }
        \pi^0 & \sqrt{2} \pi^+ \\
        \sqrt{2} \pi^- & - \pi^0
    \end{array}
\right)
\end{equation}
where the $\tau_i$'s are the Pauli matrices. The Goldstone bosons are
incorporated in a $2 \times 2$ matrix-value field $U=u_R u_L^\dag =
u_L^\dag u_R$ which is usually written in exponential form as
\begin{equation}
U(x) = \exp \left( \frac{i \phi(x)}{F_0} \right)
\end{equation}
where $F_0$ is the pion decay constant in the chiral limit. In the
pion-nucleon sector the nucleons are described by spinors denoted by
$\Psi(x)$
\begin{equation}
\Psi(x) = \left(
    \begin{array}{c}
p \\n
    \end{array}
\right) .
\end{equation}

The effective Lagrangian can be formulated as an expansion in powers
of small momenta as
\begin{equation}
\mathcal{L}_{\mathrm{eff}} = \mathcal{L}_{\pi\pi}^{(2)} + \mathcal{L}_{\pi\pi}^{(4)}
+ \mathcal{L}_{\pi N}^{(1)} + \mathcal{L}_{\pi N}^{(2)} + \mathcal{L}_{\pi N}^{(3)}
+ \mathcal{L}_{\pi N}^{(4)} + \ldots
\end{equation}
where $\mathcal{L}_{\pi \pi}^{(i)}$ and $\mathcal{L}_{\pi N}^{(i)}$
refer to the effective Lagrangian in mesonic and pion-nucleon sectors,
respectively.

The inclusion of virtual photons and leptons leads to new contact
terms and the Lagrangian is now composed of three parts
\begin{equation}
\mathcal{L}^{(i)} = \mathcal{L}_{\mathrm{strong}}^{(i)}
+ \mathcal{L}_{\mathrm{em}}^{(i)} + \mathcal{L}_{\mathrm{weak}}^{(i)} .
\end{equation}
In the mesonic sector, the explicit form of the strong Lagrangian is
given in Ref. \cite{ref:GL84, ref:GSS88}. The electromagnetic
Lagrangian is given in Ref. \cite{ref:Ure95, ref:Kne+00} and the weak
Lagrangian is given in Ref. \cite{ref:Kne+00}. In the pion-nucleon
sector, the strong and electromagnetic Lagrangians are respectively in
Ref. \cite{ref:Fet+00} and Ref. \cite{ref:MM99} up to fourth order.

Notice that the weak Lagrangian for pion-nucleon sector is needed for
the full treatment of electromagnetic corrections to weak
processes. Thus in this paper we are concerned with the construction
of the terms including both virtual photons and leptons, i.e. with
$\mathcal{L}_{\pi N,{\mathrm{weak}}}^{(i)}$.

The most general effective Lagrangian is built from pions $\phi(x)$,
nucleons $\Psi(x)$, and external scalar $s(x)$, pseudoscalar $p(x)$,
vector $v(x)$ and axial vector $a(x)$ fields. The Goldstone bosons are
described by $u_{L,R}(\phi)$ and the nucleon by the spinor $\Psi(x)$,
with the transformation properties
\begin{eqnarray}
    \begin{array}{l c l}
u_L (\phi) & \xrightarrow{G} & g_L u_L h(g,\phi)^{-1}, \nonumber \\
u_R (\phi) & \xrightarrow{G} & g_R u_R h(g,\phi)^{-1}, \nonumber \\
\Psi (x)& \xrightarrow{G} & h (g,\phi) \Psi (x), \nonumber
    \end{array}
\end{eqnarray}
\begin{equation}
g = (g_L, g_R) \in {\mathrm{SU(2)}_L} \times{\mathrm{SU(2)_R}}
\end{equation}
where $h(g,\phi)$ is the compensator defining a nonlinear realization
of $G$.

One constructs the building blocks involving external sources as
follows:
\begin{eqnarray}
u_\mu &=& i [u_R^\dag (\partial_\mu - ir_\mu)u_R + u_L^\dag(\partial_\mu
- il_\mu)u_L], \nonumber \\
f_{\mu \nu}^\pm &=& u_R^\dag F_{\mu \nu}^R u_R \pm u_L^\dag
F_{\mu \nu}^L u_L, \nonumber \\
\chi_\pm &=& u_R^\dag \chi u_L \pm u_L^\dag \chi^\dag u_R
\label{eq:building_blocks}
\end{eqnarray}
with
\begin{eqnarray}
l_\mu &=& v_\mu - a_\mu, \label{eq:normal_left_field} \\
r_\mu &=& v_\mu + a_\mu, \label{eq:normal_right_field} \\
F_{\mu \nu}^L &=& \partial_\mu l_\nu - \partial_\nu l_\mu - i[l_\mu, l_\nu],  \\
F_{\mu \nu}^R &=& \partial_\mu r_\nu - \partial_\nu r_\mu - i[r_\mu, r_\nu],  \\
\chi &=& 2 B_0 (s+i p).
\end{eqnarray}
The building blocks in Eq. (\ref{eq:building_blocks}) all transform
chirally as
\begin{equation}
X \xrightarrow{G}  h (g,\phi) X h(g,\phi)^{-1}.
\end{equation}
Covariant derivatives are defined as
\begin{eqnarray}
\nabla_\mu X  &=& \partial_\mu X + [\Gamma_\mu, X],\\
\nabla_\mu \Psi &=& \partial_\mu \Psi + \Gamma_\mu \Psi
\end{eqnarray}
for quantities transforming as $X$ above or as $\Psi$ respectively, where
\begin{equation}
\Gamma_\mu = \frac{1}{2} [u_R^\dag (\partial_\mu - ir_\mu)u_R
+ u_L^\dag (\partial_\mu - il_\mu) u_L].
\end{equation}

When virtual photons are considered, one must introduce additional
building blocks, or ``spurions'', which correspond to these virtual
photons. In the mesonic sector one works with the quark charge matrix
\begin{equation}
Q_{L,R}^{\mathrm{em}} = \frac{1}{3}\left( \begin{array}{c c}
2 & 0 \\
0 & -1
\end{array} \right)
\end{equation}
with the transformation property under chiral $SU(2)_L \times SU(2)_R$
\begin{equation}
Q_L^{\mathrm{em}} \xrightarrow{G} g_L Q_L^{\mathrm{em}} g_L^\dag, \hspace{0.2in}
Q_R^{\mathrm{em}} \xrightarrow{G}  g_R Q_L^{\mathrm{em}} g_R^\dag.
\end{equation}

In the pion-nucleon sector one uses the nucleon charge matrix instead
of the quark charge matrix \cite{ref:MS98} which is
\begin{equation} \label{eq:emchg}
Q_{L,R}^{\mathrm{em}} = \left( \begin{array}{c c}
1 & 0 \\
0 & 0
\end{array} \right) .
\end{equation}
Following Ref. \cite{ref:MM99}, one redefines the electromagnetic
spurions which refer to virtual photons as
\begin{equation} \label{eq:Q_pm}
Q_\pm^\mathrm{em} = \frac{1}{2}\{ \mathcal{Q}_L^{\mathrm{em}}
\pm \mathcal{Q}_R^{\mathrm{em}} \},
\end{equation}
where
\begin{equation}
\mathcal{Q}_{L,R}^{\mathrm{em}} = u_{L,R}^\dag Q_{L,R}^{\mathrm{em}} u_{L,R}.
\end{equation}
Then $Q_\pm^\mathrm{em}$ transforms under $SU(2)_L \times SU(2)_R$
symmetry in the same way as the other building blocks,
\begin{equation}
Q_\pm^\mathrm{em} \xrightarrow{G} h(g, \phi) Q_\pm^\mathrm{em} h(g, \phi)^{-1}.
\end{equation}

To include virtual leptons we need to introduce ``weak spurions''
related to the lepton currents. This is the procedure used by Knecht
et. al. \cite{ref:Kne+00} which we follow. Thus we take
\begin{equation} \label{eq:Qwk}
\mathcal{Q}_L^{\mathrm{wk}} = u_L^\dag Q_L^{\mathrm{wk}} u_L
\end{equation}
which transforms under chiral symmetry as
\begin{equation}
\mathcal{Q}_L^{\mathrm{wk}} \xrightarrow{G} h(g, \phi)
\mathcal{Q}_L^{\mathrm{wk}} h(g, \phi)^{-1} .
\end{equation}
In our case, since we work in $SU(2)$ symmetry whereas in
Ref. \cite{ref:Kne+00} $SU(3)$ was used, our weak spurion is
\begin{equation} \label{eq:wkchg}
Q_L^{\mathrm{wk}} = -2 \sqrt{2} G_F \left( \begin{array}{c c}
0 & V_{ud} \\
0 & 0
\end{array} \right)
\end{equation}
where $G_F$ is the Fermi coupling constant and $V_{ud}$ is the
Kobayashi-Maskawa matrix element.

Using this spurion notation we can rewrite the left handed and right
handed fields in Eq. (\ref{eq:normal_left_field}) and
Eq. (\ref{eq:normal_right_field}) as
\begin{eqnarray}
l'_\mu &=& v_\mu - a_\mu  - eQ_L^{\mathrm{em}} A_\mu
+ j_\mu^{\mathrm{wk}} Q_L^{\mathrm{wk}} + {j_\mu^{\mathrm{wk}}}^\dag
{Q_L^{\mathrm{wk}}}^\dag,  \\
r'_\mu &=& v_\mu + a_\mu - eQ_R^{\mathrm{em}} A_\mu
\end{eqnarray}
where
\begin{equation}
j_\mu^{\mathrm{wk}} = \sum_\ell \bar{\ell} \gamma_\mu
(1 - \gamma_5) \nu_\ell, \hspace{0.2in} \ell = e, \mu
\end{equation}
is the weak leptonic current and where now $v_\mu$ and $a_\mu$
represent any additional vector and axial currents that might be
presented. The contributions to processes involving purely external
photons and leptons are obtained just by replacing $l_\mu$ and $r_\mu$
by $l'_\mu$ and $r'_\mu$ in the strong part of the Lagrangian. When we
consider virtual photons and leptons however there will be additional
terms in the Lagrangian explicitly involving the weak and
electromagnetic spurions $\mathcal{Q}_L^{\mathrm{wk}}$ and
$Q_\pm^\mathrm{em}$.

The resulting Lagrangian must be invariant under the CP
transformation, so we need to know how the building blocks
transform. We find
\begin{eqnarray}
Q_\pm^\mathrm{em} && \xrightarrow{CP} (Q_\pm^\mathrm{em} )^T, \\
\mathcal{Q}_L^{\mathrm{wk}} && \xrightarrow{CP}
( - {\mathcal{Q}_L^{\mathrm{wk}}}^\dag )^T,    \\
j_\mu^{\mathrm{wk}} && \xrightarrow{CP}   {{j^{\mathrm{wk} \: \mu}}}^\dag.
\end{eqnarray}
The transformation properties of other quantities are given in
Ref. \cite{ref:Fet+00}.

\section{The leptonic Lagrangian}

All the building blocks which are necessary to construct the effective
Lagrangian have now been introduced, and we are in a position to
combine these building blocks to form invariant monomials, using a
procedure analogous to that of Ref. \cite{ref:MM99}. We consider first
the third order and then separately the fourth order effective
Lagrangian.

\subsection{Third order}

For the third order leptonic Lagrangian we have to consider terms of
the general form
\begin{equation}
e^2 j_\mu^{\mathrm{wk}} \overline{\Psi} Q_\pm^\mathrm{em}
\mathcal{Q}_L^{\mathrm{wk}} \Theta^\mu \Psi \label{eq:general_form_Lag}.
\end{equation}
The quantities $e$ and $\mathcal{Q}_L^{\mathrm{wk}}$ both count as
order $p$, so for a third order result $\Theta^\mu$ must be of order
$p^0$. Possibilities for $\Theta^\mu$ are $\gamma^\mu$, $\gamma^\mu
\gamma_5$, $\nabla^\mu$, $\gamma_5 \nabla^\mu$ and $\sigma^{\mu \nu}
\nabla_\nu$, where the $\nabla^\mu$ acts on $\Psi$ or
$\overline{\Psi}$. Covariant derivatives acting on the
$Q_\pm^\mathrm{em}$, $\mathcal{Q}_L^{\mathrm{wk}}$ or
$j_\mu^{\mathrm{wk}}$ are of order $p$ and so are not included because
they are higher order.

The possible combinations involving the product of $Q_\pm^\mathrm{em}$
and $\mathcal{Q}_L^{\mathrm{wk}}$ can be rewritten in terms of
commutators, anticommutators, and single and multiple traces as
\begin{eqnarray}
&&   [Q_\pm^\mathrm{em}, \mathcal{Q}_L^{\mathrm{wk}}], \\
&&   \{ Q_\pm^\mathrm{em}, \mathcal{Q}_L^{\mathrm{wk}} \}, \\
&&   \langle Q_\pm^\mathrm{em} \rangle \mathcal{Q}_L^{\mathrm{wk}}, \\
&&   Q_\pm^\mathrm{em} \langle \mathcal{Q}_L^{\mathrm{wk}}\rangle, \\
&&   \langle Q_\pm^\mathrm{em} \mathcal{Q}_L^{\mathrm{wk}} \rangle, \\
&&   \langle Q_\pm^\mathrm{em} \rangle \langle \mathcal{Q}_L^{\mathrm{wk}}\rangle.
\end{eqnarray}
From the definitions of $Q_-^\mathrm{em}$ and
$\mathcal{Q}_L^{\mathrm{wk}} $ in Eq. (\ref{eq:Q_pm}) and
Eq. (\ref{eq:Qwk}), and using the specific choices of
Eq. (\ref{eq:emchg}) and Eq. (\ref{eq:wkchg}), one sees that $\langle
Q_-^\mathrm{em} \rangle = \langle \mathcal{Q}_L^{\mathrm{wk}} \rangle
=0 $ so terms involving those traces do not contribute. The
anticommutator of two matrices can be written in terms of their traces
as
\begin{equation}
\{ A, B \} = A \langle B \rangle + \langle A \rangle B
+ \langle AB \rangle - \langle A \rangle \langle B \rangle
\end{equation}
which leads to
\begin{eqnarray}
\left\{ Q_+^\mathrm{em}, \mathcal{Q}_L^{\mathrm{wk}} \right\} &=&
\langle Q_+^\mathrm{em} \rangle \mathcal{Q}_L^{\mathrm{wk}}
+ \langle Q_+^\mathrm{em} \mathcal{Q}_L^{\mathrm{wk}} \rangle, \\
\left\{ Q_-^\mathrm{em}, \mathcal{Q}_L^{\mathrm{wk}} \right\} &=&
\langle Q_-^\mathrm{em} \mathcal{Q}_L^{\mathrm{wk}} \rangle
\end{eqnarray}
and allows us to eliminate two of the possibilities as not independent.
Thus all possible independent chiral invariant terms can be written as
\begin{equation}
e^2 \overline{\Psi} j_\mu^{\mathrm{wk}} \left\{
\begin{array}{c}
[Q_\pm^\mathrm{em}, \mathcal{Q}_L^{\mathrm{wk}}] \\
\langle Q_+^\mathrm{em} \rangle \mathcal{Q}_L^{\mathrm{wk}} \\
\langle Q_\pm^\mathrm{em} \mathcal{Q}_L^{\mathrm{wk}} \rangle
\end{array}
\right\} \Theta^\mu \Psi + {\mathrm{h.c.}}. \label{eq:poss_chiral_inv}
\end{equation}

However, the leptonic Lagrangian must be not only chiral but also CP
invariant. Therefore we have to consider the CP transformation
properties of each term, and construct CP invariant combinations. Once
this is done we can look again at the possibilities for $ \Theta^\mu$.
The term $\sigma^{\mu \nu} \nabla_\nu$ involves a $\nabla_\mu
\gamma^\mu$ so we can use the equation of motion (EOM) to express it
in terms of the other possibilities. Likewise the CP invariant
combinations involving $\nabla^\mu$ and $\gamma_5 \nabla^\mu$ can be
shown to be higher order or not independent of the other terms using
total derivative \cite{ref:FS96} or EOM arguments. Thus the only
remaining independent possibilities for $ \Theta^\mu$ are $\gamma^\mu$
and $\gamma^\mu \gamma_5$.  Thus our final result for the third order
Lagrangian can be written in the form
\begin{equation}
\mathcal{L}_{\pi N,{\mathrm{wk}}}^{(3)} = e^2 F_0^2 \sum_i n_i \overline{\Psi}
\mathcal{O}_{i,\mathrm{wk}}^{(3)} \Psi + {\mathrm{h.c.}}
\end{equation}
where $n_i$ are the third order low energy coupling constants (LECs)
and the $\mathcal{O}_{i, \mathrm{wk}}^{(3)}$ are shown in
Table. \ref{tab:O_wk3}.

\begin{table}[tbh!]
\caption{The monomials $\mathcal{O}_{i,{\mathrm{wk}}}^{(3)}$ of third
  order for the leptonic Lagrangian. $\widetilde{Q}^\mathrm{em}_\pm =
  Q_\pm^\mathrm{em} - \frac{1}{2} \langle Q_\pm^\mathrm{em} \rangle $
  is the traceless part of the operator $Q_\pm^\mathrm{em}$.}
\label{tab:O_wk3}
\begin{center}
\begin{tabular}{|c|l|}
\hline
i & $\mathcal{O}_{\mathrm{wk}}(e^2 p)$ \\
\hline
1 & $ \gamma^\mu j_\mu^{\mathrm{wk}} \langle \widetilde{Q}^\mathrm{em}_+ \mathcal{Q}_L^\mathrm{wk} \rangle$  \\
2 & $ \gamma^\mu j_\mu^{\mathrm{wk}}  \langle Q_+^\mathrm{em}\rangle \mathcal{Q}_L^\mathrm{wk} $  \\
3 & $ \gamma^\mu j_\mu^{\mathrm{wk}} [ \widetilde{Q}^\mathrm{em}_+ , \mathcal{Q}_L^\mathrm{wk}]$  \\
4 & $ \gamma^\mu j_\mu^{\mathrm{wk}} \langle \widetilde{Q}^\mathrm{em}_- \mathcal{Q}_L^\mathrm{wk} \rangle $  \\
5 & $ \gamma^\mu j_\mu^{\mathrm{wk}} [ \widetilde{Q}^\mathrm{em}_- , \mathcal{Q}_L^\mathrm{wk}] $  \\
6 & $ \gamma^\mu \gamma_5 j_\mu^{\mathrm{wk}}  \langle \widetilde{Q}^\mathrm{em}_+ \mathcal{Q}_L^\mathrm{wk} \rangle$  \\
7 & $ \gamma^\mu \gamma_5 j_\mu^{\mathrm{wk}}  \langle Q_+^\mathrm{em} \rangle \mathcal{Q}_L^\mathrm{wk}$  \\
8 & $ \gamma^\mu \gamma_5 j_\mu^{\mathrm{wk}} [ \widetilde{Q}^\mathrm{em}_+ , \mathcal{Q}_L^\mathrm{wk}]$  \\
9 & $ \gamma^\mu \gamma_5 j_\mu^{\mathrm{wk}} \langle \widetilde{Q}^\mathrm{em}_- \mathcal{Q}_L^\mathrm{wk} \rangle$  \\
10 & $\gamma^\mu \gamma_5 j_\mu^{\mathrm{wk}} [ \widetilde{Q}^\mathrm{em}_- , \mathcal{Q}_L^\mathrm{wk}]$  \\
\hline
\end{tabular}
\end{center}
\end{table}

\begin{table}
\caption{The monomials $\mathcal{O}_{i,{\mathrm{wk}}}(e^2 p^2)$ of
  fourth order for the leptonic Lagrangian, in the limit of no pions.  }
\label{tab:O^4wk}
\begin{center}
\begin{tabular}{|c|l|c|l|}
\hline
i & $\mathcal{O}_{\mathrm{wk}}(e^2 p^2)$ & i & $\mathcal{O}_{\mathrm{wk}}(e^2 p^2)$\\
\hline
1 & $i (\nabla^\mu j_\mu^\mathrm{wk})\langle \widetilde{Q}^\mathrm{em}_+ \mathcal{Q}_L^\mathrm{wk} \rangle $   &
11 & $ \sigma^{\mu \nu} \gamma_5 (\nabla_\nu j_\mu^\mathrm{wk}) \langle \widetilde{Q}^\mathrm{em}_+ \mathcal{Q}_L^\mathrm{wk}\rangle$   \\
2 & $i (\nabla^\mu j_\mu^\mathrm{wk})  \langle Q_+^\mathrm{em} \rangle \mathcal{Q}_L^\mathrm{wk} $   &
12 & $ \sigma^{\mu \nu} \gamma_5 (\nabla_\nu j_\mu^\mathrm{wk}) \langle Q_+^\mathrm{em} \rangle \mathcal{Q}_L^\mathrm{wk} $  \\
3 & $i (\nabla^\mu j_\mu^\mathrm{wk}) \left[ \widetilde{Q}^\mathrm{em}_+, \mathcal{Q}_L^\mathrm{wk} \right]$   &
13 & $ \sigma^{\mu \nu} \gamma_5 (\nabla_\nu j_\mu^\mathrm{wk}) \left[ \widetilde{Q}^\mathrm{em}_+, \mathcal{Q}_L^\mathrm{wk}  \right] $  \\
4 & $i (\nabla^\mu j_\mu^\mathrm{wk})\langle \widetilde{Q}^\mathrm{em}_- \mathcal{Q}_L^\mathrm{wk} \rangle$   &
14 & $ \sigma^{\mu \nu} \gamma_5 (\nabla_\nu j_\mu^\mathrm{wk}) \langle \widetilde{Q}^\mathrm{em}_- \mathcal{Q}_L^\mathrm{wk} \rangle$   \\
5 & $i (\nabla^\mu j_\mu^\mathrm{wk}) \left[ \widetilde{Q}^\mathrm{em}_-,\mathcal{Q}_L^\mathrm{wk} \right]$   &
15 &  $ \sigma^{\mu \nu} \gamma_5 (\nabla_\nu j_\mu^\mathrm{wk}) \left[ \widetilde{Q}^\mathrm{em}_-, \mathcal{Q}_L^\mathrm{wk}\right]$  \\
6 & $ \sigma^{\mu \nu} (\nabla_\nu j_\mu^\mathrm{wk}) \langle \widetilde{Q}^\mathrm{em}_+ \mathcal{Q}_L^\mathrm{wk} \rangle$   &
16 & $ i \gamma_5 (\nabla^\mu j_\mu^\mathrm{wk})\langle \widetilde{Q}^\mathrm{em}_+ \mathcal{Q}_L^\mathrm{wk} \rangle$   \\
7 & $ \sigma^{\mu \nu} (\nabla_\nu j_\mu^\mathrm{wk}) \langle Q_+^\mathrm{em} \rangle \mathcal{Q}_L^\mathrm{wk}$   &
17 & $ i \gamma_5 (\nabla^\mu j_\mu^\mathrm{wk})  \langle Q_+^\mathrm{em} \rangle \mathcal{Q}_L^\mathrm{wk}$   \\
8 & $ \sigma^{\mu \nu} (\nabla_\nu j_\mu^\mathrm{wk}) \left[ \widetilde{Q}^\mathrm{em}_+ ,\mathcal{Q}_L^\mathrm{wk} \right] $   &
18 & $ i \gamma_5 (\nabla^\mu j_\mu^\mathrm{wk}) \left[ \widetilde{Q}^\mathrm{em}_+, \mathcal{Q}_L^\mathrm{wk} \right]$   \\
9 & $ \sigma^{\mu \nu} (\nabla_\nu j_\mu^\mathrm{wk}) \langle \widetilde{Q}^\mathrm{em}_- \mathcal{Q}_L^\mathrm{wk} \rangle$   &
19 & $ i \gamma_5 (\nabla^\mu j_\mu^\mathrm{wk})\langle \widetilde{Q}^\mathrm{em}_- \mathcal{Q}_L^\mathrm{wk}\rangle$   \\
10 & $ \sigma^{\mu \nu} (\nabla_\nu j_\mu^\mathrm{wk}) \left[ \widetilde{Q}^\mathrm{em}_- ,\mathcal{Q}_L^\mathrm{wk} \right]$   &
20 & $ i \gamma_5 (\nabla^\mu j_\mu^\mathrm{wk}) \left[ \widetilde{Q}^\mathrm{em}_-, \mathcal{Q}_L^\mathrm{wk} \right]$   \\
\hline
\end{tabular}
\end{center}
\end{table}

\subsection{Fourth order}

Consider now the fourth order Lagrangian.  By using the same methods
as in the previous section, we find that the general structure of the
fourth order Lagrangian is the same as for the third order, as shown
in Eq. (\ref{eq:poss_chiral_inv}). However now $\Theta^\mu$ must be of
order $p$.  The building blocks which are of order $p$ which are at
our disposal are $u_\nu$ and $\nabla_\nu$, where $\nabla_\nu$ acts on
one of the internal quantities, $j_\mu^\mathrm{wk}$,
$\mathcal{Q}_L^\mathrm{wk}$, or $Q_\pm^\mathrm{em}$. The residual part
of $\Theta^\mu$, after extracting one of these building blocks, takes
the form $\Theta^{\mu\nu}$ and is of order $p^0$.

Possibilities for $\Theta^{\mu\nu}$ which do not include any
additional derivatives are $g^{\mu\nu}$, $g^{\mu\nu} \gamma_5$,
$\sigma^{\mu \nu}$, and $\epsilon^{\mu\nu\alpha\beta}
\sigma_{\alpha\beta}$. There is also a second set of possibilities
which involve derivatives acting on $\Psi$, which, recall, are of
order $p^0$. These include $\nabla^\mu \nabla^\nu$, $\nabla^\mu
\nabla^\nu \gamma_5$, $\gamma^\mu \nabla^\nu$, $\gamma^\nu
\nabla^\mu$, $\gamma^\mu \gamma_5 \nabla^\nu$, $\gamma^\nu \gamma_5
\nabla^\mu$, $\epsilon^{\mu\nu\alpha\beta} \gamma_\alpha \nabla_\beta$,
and $\epsilon^{\mu\nu\alpha\beta} \gamma_\alpha \gamma_5 \nabla_\beta$.

The next step is to construct CP invariant and Hermitian combinations
of these possibilities. It is then straightforward, but somewhat
tedious, to show using total derivative and EOM arguments, just as
done for the third order case, that all of the possibilities involving
derivatives acting on $\Psi$ can be expressed in terms of the first
set of terms, or are of higher order.

The final set of terms can then be listed and consists of terms with
one of the four structures without derivatives, one of the five
combinations of $\mathcal{Q}_L^\mathrm{wk}$ and $Q_\pm^\mathrm{em}$
and either $u_\nu$ (taken in all possible orders) or $\nabla_\nu$
acting on $j_\mu^\mathrm{wk}$, $\mathcal{Q}_L^\mathrm{wk}$, or
$Q_\pm^\mathrm{em}$. Since there are so many terms we will give
explicitly only the set most useful for standard weak interactions,
namely those without extra pions. In that limit $u_\nu$ vanishes as do
the derivatives acting on $\mathcal{Q}_L^\mathrm{wk}$ and
$Q_\pm^\mathrm{em}$. Thus all terms involve $\nabla_\nu j_\mu^\mathrm{wk}$,
or actually just $\partial_\nu j_\mu^\mathrm{wk}$.

Thus in this limit of no pions the fourth order effective Lagrangian
is given by
\begin{equation}
\mathcal{L}_{\pi N,{\mathrm{wk}}}^{(4)} = e^2 F_0^2 \sum_i s_i \overline{\Psi}
\mathcal{O}_{i,\mathrm{wk}}^{(4)} \Psi + h.c.
\end{equation}
where $s_i$ are the fourth order low-energy coupling constants. The
$\mathcal{O}_{i,\mathrm{wk}}^{(4)}$ monomials are tabulated in Table
\ref{tab:O^4wk}. Note that in the table we have rewritten
$\epsilon^{\mu\nu\alpha\beta} \sigma_{\alpha\beta}$ as
$\sigma^{\mu\nu} \gamma_5$ using standard identities.

\section{Acknowledgments}

This work was supported in part by the Thailand Research Fund under
Grant No. PHD/0221/2545 and in part by the Natural Sciences and
Engineering Research Council of Canada.

%\bibliography{bib}

%

\end{document}